\documentclass[aps,prd]{revtex4}

\usepackage{graphicx} 
\usepackage[scriptsize,hang,flushleft]{caption}
\usepackage[scriptsize,hang]{subfigure}
\usepackage{color}
\usepackage{placeins}

\usepackage{amsmath,amsfonts,bbm,dsfont,mathrsfs}

\newcommand{\half}{\frac{1}{2}}
\newcommand{\dd}{{\mathrm{d}}}

\begin{document}

\title{Motion of Charged Particles around a\\ Scalarized Black Hole in Kaluza-Klein Theory}

\author{Saskia Grunau}
\email[]{saskia.grunau@uni-oldenburg.de}
\author{Matthias Kruse}
\email[]{matthias.kruse@uni-oldenburg.de}
\affiliation{Institut f\"ur Physik, Universit\"at Oldenburg, D--26111 Oldenburg, Germany}

\date{\today}

\begin{abstract}
In this article we study the motion of charged particles in the spacetime of a charged rotating scalarized black hole, namely the rotating dyonic black hole in Kaluza-Klein theory found by Rasheed. We derive the equations of motion in the extremal case and present their analytical solutions in terns of elliptic and hyperelliptic functions. Furthermore, we analyze the particle motion with the help of parametric diagrams and effective potentials. 
\end{abstract}

\maketitle

\section{Introduction}

Recently scalarized black holes gained a lot of attention in the literature. In general relativity black holes possess no scalar hair, associated with a single real scalar field. They are uniquely determined by their mass, angular momentum and charge if gravity is coupled to a Maxwell field \cite{Chrusciel:2012jk}. However, there are possibilities to bypass the no-hair theorem and allow scalar hair on black holes \cite{Herdeiro:2015waa, Cardoso:2016ryw}. In string theory the scalar field corresponds to a dilaton. Coupling the dilaton to the Lagrangian of the electromagnetic field yields charged dilatonic black hole solutions  \cite{Garfinkle:1990qj}, \cite{Kleihaus:2003df}, \cite{Gibbons:1987ps}. In the case of Kaluza-Klein coupling, first a new spatial dimension is added to an existing black hole solution, then a boost is performed. The charged dilatonic black hole solution can now be obtained by a dimensional reduction, see  \cite{Maison:1979kx}--\cite{Kunz:2006jd}.

Chodos et al. applied this method \cite{Chodos:1980df} to the Schwarzschild spacetime and obtained the static electrically charged Einstein-Maxwell-dilaton black hole. The corresponding rotating charged solution was found in \cite{Frolov:1987rj,Horne:1992zy}, where the Kerr solution was used as a starting point for the method. Later Rasheed \cite{Rasheed:1995zv} applied two boosts and a rotation to find a  rotating, dilatonic black hole in  Kaluza-Klein theory which is  electrically and magnetically charged.

The dilaton can also be coupled to a curvature invariant, for example a Gau{\ss}-Bonnet term \cite{Kanti:1995vq}--\cite{Kokkotas:2017ymc}. Lately, there is a growing interest in more general coupling functions, since these lead to spontaneously scalarized black holes \cite{Doneva:2017bvd}--\cite{Konoplya:2019goy}.\\

A powerful approach to gain insight to the structure of a spacetime is to study the motion of test particles. Black holes in different theories can be explored with the analysis of the orbits of test particles and light. When deriving the equations of motion, the Hamilton-Jacobi formalism is a very efficient method. For the Kerr spacetime, the separability of the Hamilton-Jacobi equation was shown by Carter \cite{Carter:1968ks}. In the four-dimensional Kerr spacetime, the equations of motion can be solved analytically with the help of elliptic functions. However, in higher dimensions or when the cosmological constants is taken into account, the equations become often more complicated and  hyperelliptic functions are needed in order to solve the equations of motion analytically, see e.g. \cite{Kraniotis:2003ig}--\cite{Hackmann:2010zz}.

Most scalarized  black hole solutions exist in numerical form. In this article we want to focus on an analytical solution and therefore we study the spacetime of the extremal rotating dyonic black hole in Kaluza-Klein theory found by Rasheed \cite{Rasheed:1995zv}. This solution has some surprising features, which make it interesting to study. Rasheed found that the gyromagnetic and gyroelectric ratios can become arbitrarily large. Furthermore he investigated the properties of the extremal case. The angular velocity of the event horizon can be zero while the black hole still has non-zero ADM angular momentum. Also, in certain configurations angular momentum can be added, while keeping the mass and charges fixed, and the solution will remain extremal. In Einstein-Maxwell theory this would lead to a naked singularity.

We will investigate the motion of massless particles and charged particles in the black hole spacetime of Rasheed. A subclass of spacetimes with electric and dilatonic charge only, the Kerr-Kaluza-Klein black hole, was considered in \cite{Aliev:2013jya}. The authors showed that the Hamilton-Jacobi equation separates completely for massless particles. Additionally they studied the motion uncharged particles in the equatorial plane. In \cite{Cunha:2018uzc} black holes without $\mathbb{Z}_2$ symmetry are considered, which includes the dyonic black hole found by Rasheed \cite{Rasheed:1995zv}. It was shown that the Hamilton-Jacobi equation is fully separable for massless particles.\\

The present article is structured as follows. First we will introduce the black hole spacetime by Rasheed \cite{Rasheed:1995zv} and review some of its properties. It can be shown that the Hamilton-Jacobi equation for particles moving around an extremal Rashed black hole separates in three cases: massless particles, charged particles around a non-rotating Rasheed black hole, charged particles in the equatorial plane of the rotating Rasheed black hole. We will analyze the motion of charged particles in the equatorial plane in detail. Then we will present the analytical solutions of the equations of motion and plot some examples of the orbits.

\section{The rotating dyonic black hole in Kaluza-Klein theory}

In 1995 D. Rasheed \cite{Rasheed:1995zv} presented a rotating dyonic black hole in Kaluza-Klein theory. It can be obtained from the Kerr solution by adding a spatial dimension and then applying two boosts and a rotation. A dimensional reduction leads to an electrically and magnetically charged rotating black hole with a scalar dilaton field. The metric is
\begin{equation}
  \dd s^2_{(4)} = -\frac{f^2}{\sqrt{AB}}\left(\dd t+{\omega^0}_\phi \dd\phi\right)^2 + \frac{\sqrt{AB}}{\Delta}\dd r^2 + \sqrt{AB}\dd\theta^2 + \frac{\Delta\sqrt{AB}}{f^2}\sin^2\theta \dd\phi^2
\label{eqn:ds4}
\end{equation}
and the electromagnetic vector potential is given by
\begin{equation}
2A_\mu \dd x^\mu = \frac{C}{B}\dd t + \left({\omega^5}_\phi + \frac{C}{B}{\omega^0}_\phi\right)\dd\phi \, .
\end{equation}
The dilaton field $\sigma$ can be read of from the five-dimensional metric in \cite{Rasheed:1995zv} which gives
\begin{equation}
 \exp \left( \frac{4}{\sqrt{3}}\sigma \right) = \frac{B}{A}
\end{equation}
so that
\begin{equation}
\sigma = \frac{\sqrt{3}}{4} \ln \left( \frac{B}{A} \right)\, .
\end{equation}
The functions in the metric, the vector potential and the dilaton field are
\begin{align}
  A &= \left(r-\Sigma /\sqrt{3}\right)^2 - \frac{2P^2\Sigma}{\Sigma - M\sqrt{3}} + a^2\cos^2\!\theta + \frac{2JPQ\cos\theta}{\left(M+\Sigma/\sqrt{3}\right)^2-Q^2} \, ,
  \\
  B &= \left(r+\Sigma /\sqrt{3}\right)^2 - \frac{2Q^2\Sigma}{\Sigma + M\sqrt{3}} + a^2\cos^2\!\theta - \frac{2JPQ\cos\theta}{\left(M-\Sigma/\sqrt{3}\right)^2-P^2} \, ,
  \\
  C &= 2 Q \left(r-\Sigma /\sqrt{3}\right) - \frac{2PJ\cos\theta\left(M+\Sigma /\sqrt{3}\right) }{\left(M-\Sigma /\sqrt{3}\right)^2-P^2} \, ,
  \\
  {\omega^0}_\phi &= \frac{2J\sin^2\!\theta}{ f^2}\left[r-M + \frac{\left(M^2+\Sigma^2-P^2-Q^2\right)\left(M+\Sigma /\sqrt{3}\right)}{\left(M+\Sigma /\sqrt{3}\right)^2-Q^2}\right] \, ,
  \\
  {\omega^5}_\phi &= \frac{2P\Delta}{ f^2}\cos\theta - \frac{2QJ\sin^2\!\theta \left[r\left(M - \Sigma/\sqrt{3}\right) + M\Sigma/\sqrt{3} + \Sigma^2-P^2-Q^2\right] }{ f^2\left[\left(M+\Sigma/\sqrt{3}\right)^2-Q^2\right]}
\end{align}
and
\begin{align}
  \Delta = r^2 - 2Mr + P^2 + Q^2 - \Sigma^2 + a^2 \, , 
  \\
  f^2 = r^2 - 2Mr + P^2 + Q^2 - \Sigma^2 + a^2\cos^2\!\theta \, .
\end{align}
The parameter $a$ is related to $J$ via
\begin{equation}
J^2 = a^2\frac{\left[\left(M+\Sigma/\sqrt{3}\right)^2-Q^2\right]
\left[\left(M-\Sigma/\sqrt{3}\right)^2-P^2\right]}{ M^2+\Sigma^2-P^2-Q^2} \, 
\end{equation}
and the charges satisfy the equation
\begin{equation}
\frac{Q^2}{\Sigma+M\sqrt{3}} + \frac{P^2}{\Sigma-M\sqrt{3}} = \frac{2\Sigma}{ 3} \, .
\end{equation}
So the rotating dyonic black hole solution depends on four parameters $M$, $J$, $Q$ and $P$, where $M$ is the mass, $J$ is the angular momentum, $Q$ and $P$ are the electric and magnetic charges and $\Sigma$ is the dilaton charge. These parameters are related to the mass $M_K$ of the Kerr solution by
\begin{equation}
  M_K^2=M^2+\Sigma^2-P^2-Q^2 \, .
\end{equation}
For some purposes it is useful to express the quantities in terms of the boost parameters $\alpha$ and $\beta$, the Kerr mass $M_K$ and the rotation parameter $a$ by
\begin{align}
  M &= \frac{M_K\left(1+\cosh^2\!\alpha\cosh^2\!\beta\right)\cosh\alpha}{\sqrt{1+\sinh^2\!\alpha\cosh^2\!\beta}} \, ,
  \\
  \Sigma &= \frac{\sqrt{3}M_K\cosh\alpha\left(1-\cosh^2\!\beta+\sinh^2\!\alpha\cosh^2\!\beta\right)}{2\sqrt{1+\sinh^2\!\alpha\cosh^2\!\beta}} \, ,
  \\
  Q &= M_K\sinh\alpha\sqrt{1+\sinh^2\!\alpha\cosh^2\!\beta} \, ,
  \\
  P &= \frac{M_K\sinh\beta\cosh\beta}{\sqrt{1+\sinh^2\!\alpha\cosh^2\!\beta}}\, ,
  \\
  J &= aM_K\cosh\beta\sqrt{1+\sinh^2\!\alpha\cosh^2\!\beta} \, .
\end{align}

\subsection{Extremal solutions}

\begin{figure}[h]
  \centering
 \subfigure[Part of the surface with positive parameters]{
  \includegraphics[width=0.4\linewidth]{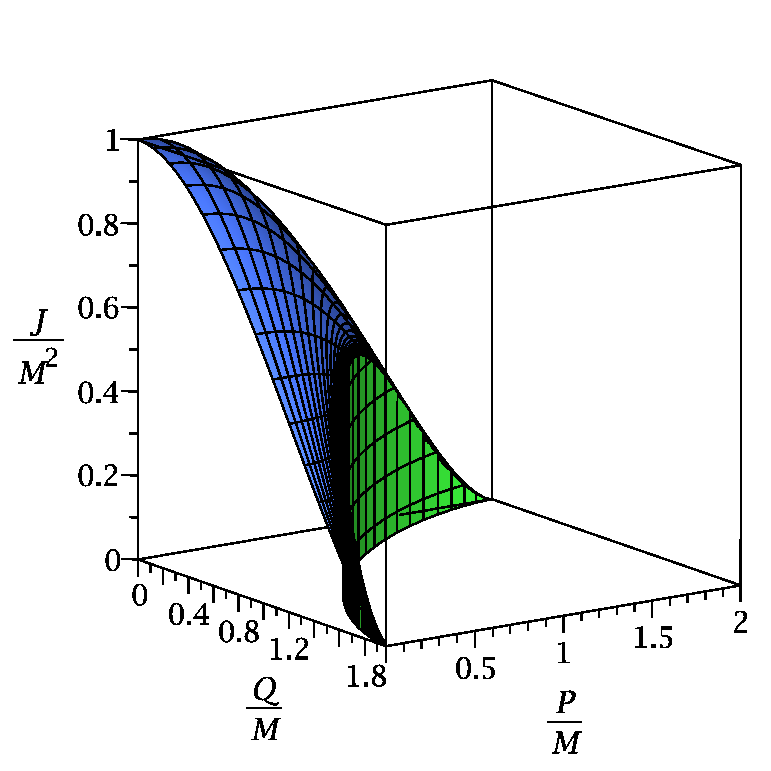}
 }
 \subfigure[Complete surface]{
  \includegraphics[width=0.4\linewidth]{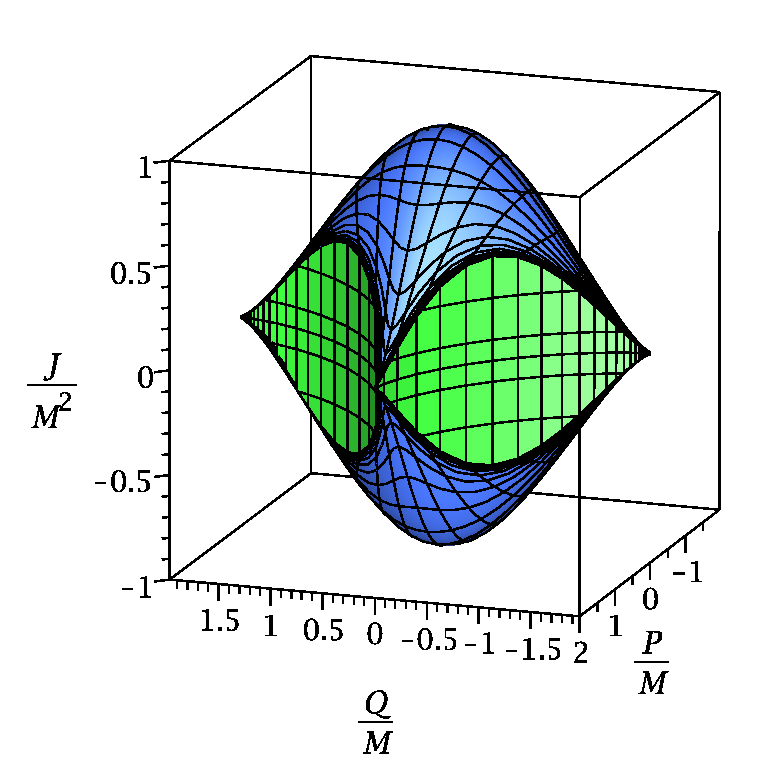}
 }
  \caption{Surface of extremal rotating dyonic solutions in Kaluza-Klein theory.}
  \label{pic:extremesurface}
\end{figure}

The two horizons of the rotating dyonic black hole are given by $\Delta = 0$. The horizons are present for
\begin{equation}
 M^2 \geq P^2+Q^2+a^2-\Sigma^2 \, .
\end{equation}
This yields $M_K^2 \geq a^2$, which is the same condition as for the original Kerr spacetime to have horizons. It is interesting to the study the extremal case with a single degenerate horizon. In Einstein-Maxwell theory the surface of extreme solutions, formed by the scaled global charges $\frac{J}{M^2}$, $\frac{Q}{M}$ and  $\frac{P}{M}$, is a sphere. In Kaluza-Klein theory this surface is not smooth but made up of two parts as shown in figure \ref{pic:extremesurface} (see also \cite{Rasheed:1995zv}).

The first surface (blue in figure \ref{pic:extremesurface}) emerges from the condition for a single degenerate horizon $ M^2 = P^2+Q^2+a^2-\Sigma^2$, which leads to $M_K=a^2$ and therefore boosting the extremal Kerr solution will lead to the blue surface.

The second surface (green in figure \ref{pic:extremesurface}) represents the special case $M^2+\Sigma^2=P^2+Q^2$. Then $a=0$ and  $M_K=0$ but $\frac{a}{M_K}\leq 1$, so that the angular momentum $J$ may be non-zero. This can be seen by taking the extremal limit $\beta\rightarrow\infty$ of the boosted solutions, thus
\begin{align}
  \frac{J}{M^2} &= \frac{4a\sinh^3\!\alpha}{M_K\cosh^6\!\alpha} \, ,\\
  \frac{P}{M} &= \frac{2}{\cosh^3\!\alpha} \, ,\\
  \frac{Q}{M} &= \frac{2\sinh^3\!\alpha}{\cosh^3\!\alpha} \, .
\end{align}
Then we have
\begin{equation}
  \left(\frac{P}{M}\right)^\frac{2}{3} + \left(\frac{Q}{M}\right)^\frac{2}{3} = 2 ^\frac{2}{3} \quad \text{and} \quad J\leq PQ\, .
\end{equation}
So any non-rotating dyonic solution with fixed $M,P,Q$ on this part of the surface can be given angular momentum $J\leq PQ$ and it will still remain extremal. In Einstein-Maxwell adding angular momentum to an extremal black hole would lead to a naked singularity.\\

In this article we will concentrate on the special case 
\begin{equation}
  P=Q=\frac{M}{\sqrt{2}} \quad \text{and} \quad J\leq PQ = \frac{M^2}{2}\, ,
\end{equation}
where the metric is
\begin{equation}
  \dd s^2 = -\frac{\left(r-M\right)^2}{\sqrt{r^4-4J^2\cos^2\!\theta}}\left(\dd t-\frac{2J\sin^2\!\theta}{ r-M}\dd\phi\right)^2 + \frac{\sqrt{r^4-4J^2\cos^2\!\theta}}{\left(r-M\right)^2}\dd r^2 + \sqrt{r^4-4J^2\cos^2\!\theta} \ ( \dd\theta^2 + \sin^2\!\theta \ \dd\phi^2),
  \label{eqn:extremalmetric}
\end{equation}
and the non-zero components of the electromagnetic vector potential are
\begin{align}
 A_t &= \frac{Mr-2J\cos\theta}{\sqrt{2}\ (r^2-2J\cos\theta) } \, , \\
 A_\phi &=  \frac{M}{\sqrt{2}}\cos\theta - \frac{\sqrt{2} \ J\sin^2\!\theta}{r-M} + \frac{\sqrt{2} \ J\sin^2\!\theta  (Mr-2J\cos\theta) }{(r-M)(r^2-2J\cos\theta)} \, .
\label{eqn:extremalAtAphi}
\end{align}
The metric possesses a singularity at $r_S=\sqrt{2 \lvert J\cos\theta \rvert}$, which is hidden behind an event horizon at $r_H=M$ if $J< \frac{M^2}{2}$. As for all solutions on the green surface, there is no ergoregion.

In the extremal case the dilaton field is
\begin{equation}
\sigma = \frac{\sqrt{3}}{4} \ln \left( \frac{r^2+2J\cos\theta}{r^2-2J\cos\theta} \right)\, ,
\label{eqn:extr-dfield}
\end{equation}
which means that $\sigma =0$ for $J=0$ and also $\sigma =0$ in the equatorial plane $\theta=\frac{\pi}{2}$.

\subsection{Equations of motion for electrically charged particles in the extremal case}
\label{sec:EQM}

Let us investigate the motion of electrically charged particles around the extremal dyonic black hole described by the equation \eqref{eqn:extremalmetric}. As described in \cite{Maki:1992up, Pris:1995, Rahaman:2003wv} we will take into account the effects of the dilaton field. The Hamiltonian for a charged particle with the electric charge $q$ is
\begin{equation}
 \mathcal{H}=\half\mathrm{e}^{-\beta \sigma}  g^{\mu\nu} \left( p_\mu - qA_\mu \right) \left( p_\nu - qA_\nu \right)
\end{equation}
where the parameter $\beta$ describes the coupling to the dilaton field $\sigma$. The mass shell condition is in this case
\begin{equation}
 g^{\mu\nu} \left( p_\mu - qA_\mu \right) \left( p_\nu - qA_\nu \right) + \mathrm{e}^{\beta \sigma}m^2 = 0 \, .
\end{equation}
The Hamilton-Jacobi equation is
\begin{equation}
  \mathcal{H} + \frac{\partial S}{\partial \lambda} =0 
 \label{eqn:hjeq}
\end{equation}
with $p_\mu = \frac{\partial S}{\partial x^\mu}$  and an affine parameter $\lambda$ along the trajectory of the test particle. To solve equation \eqref{eqn:hjeq} we use the following ansatz for the action 
\begin{equation}
 S= \half\delta\mathrm{e}^{\beta \sigma}\lambda - Et + L\phi + S_\theta(\theta) + S_r(r)
\label{eqn:action}
\end{equation}
where $E$ is the energy and $L$ is the angular momentum of the test particle. The parameter $\delta = m^2$ describes the particle's mass and is $1$ for particles and $0$ for light. With this ansatz the Hamilton-Jacobi equation becomes
\begin{equation}
   \mathrm{e}^{2\beta \sigma}\delta + g^{tt} (-E-qA_t)^2 + 2g^{t\phi}(-E-qA_t)(L-qA_\phi) + g^{\phi\phi} (L-qA_\phi)^2 + g^{\theta\theta } \left(  \frac{\partial S_\theta(\theta)}{\partial \theta} \right)^2 + g^{rr} \left(  \frac{\partial S_r(r)}{\partial r} \right)^2 = 0
\end{equation}
Inserting the metric \eqref{eqn:extremalmetric}, the vector potential \eqref{eqn:extremalAtAphi} and the dilaton field \eqref{eqn:extr-dfield} yields
\begin{align}
 &\delta \left( \frac{r^2+2J\cos\theta}{r^2-2J\cos\theta} \right) ^{\beta\sqrt{3}/2}\sqrt{r^4-4J^2\cos^2\!\theta} + \frac{4 J^2 -r^4}{(r-M)^2} \left( -\frac{q (Mr-2J\cos\theta) }{\sqrt{2}(r^2-2J \cos\theta)}-E\right)^2 \nonumber\\
 &+\frac{4J}{r-M}\left( -\frac{q (Mr-2J\cos\theta)}{\sqrt{2}(r^2-2J \cos\theta)}-E\right) \left(L -q \left( \frac{M}{\sqrt{2}} \cos\theta - \frac{\sqrt{2}J\sin^2\!\theta}{r-M} +\frac{\sqrt{2}J\sin^2\!\theta (Mr-2J\cos\theta)}{(r-M)(r^2-2J\cos\theta)} \right)\right) \nonumber\\
 &+\frac{1}{\sin^2\!\theta}\left(L -q \left( \frac{M}{\sqrt{2}} \cos\theta - \frac{\sqrt{2}J\sin^2\!\theta}{r-M} +\frac{\sqrt{2}J\sin^2\!\theta (Mr-2J\cos\theta)}{(r-M)(r^2-2J\cos\theta)} \right)\right)^2 \nonumber\\
 &+  \left(  \frac{\partial S_\theta(\theta)}{\partial \theta} \right)^2 +(r-M)^2 \left(  \frac{\partial S_r(r)}{\partial r} \right)^2 = 0 \, .
\label{eqn:HJD-general}
\end{align}
Now we have to separate the above equation to get (the derivatives of) the unknown functions $S_\theta(\theta)$ and $S_r(r)$. This is possible in three cases:
\begin{enumerate}
 \item For photons with $\delta=0$ and $q=0$, equation \eqref{eqn:HJD-general} simplifies to
	\begin{equation}
	 \frac{(4J^2-r^4)E^2}{(r-M)^2} - \frac{4JEL}{r-M} + \frac{L^2}{\sin^2\!\theta} + \left(  \frac{\partial S_\theta(\theta)}{\partial \theta} \right)^2 + (r-M)^2 \left(  \frac{\partial S_r(r)}{\partial r} \right)^2 = 0 \, .
	\end{equation}
 \item For a non-rotating black hole with $J=0$, equation \eqref{eqn:HJD-general} yields
	\begin{equation}
	\delta r^2 - \frac{r^4}{(r-M)^2}\left(-\frac{Mq}{\sqrt{2}r}-E\right)^2 + \frac{1}{\sin^2\!\theta} \left( L-\frac{Mq}{\sqrt{2}}\cos\theta \right)^2 + \left(  \frac{\partial S_\theta(\theta)}{\partial \theta} \right)^2 + (r-M)^2 \left(  \frac{\partial S_r(r)}{\partial r} \right)^2 = 0 \, .
	\end{equation}
 \item For charged particles in the equatorial plane  $\theta = \frac{\pi}{2}$, equation \eqref{eqn:HJD-general} becomes
	\begin{equation}
	 \delta r^2 -\frac{\left(r^4-4J^2\right)\left(qM\sqrt{2}+2Er\right)^2}{4r^2(r-M)^2} - \frac{2J\left(qM\sqrt{2}+2Er\right)\left(Jq\sqrt{2}+Lr\right)}{(r-M)r^2}
         +\frac{\left(Jq\sqrt{2}+Lr\right)^2}{r^2} + (r-M)^2 \left(  \frac{\partial S_r(r)}{\partial r} \right)^2 = 0 \, .
	  \label{eqn:HJD4-equa}
	\end{equation}
\end{enumerate}
So far we have described the particle motion in the four-dimensional Rasheed spacetime, which respresents a scalarized black hole. Since most scalarized black holes are given as a numerical solution, it is interesting to have some insight on the analytical side.

On the other hand, one has to take into account that the four-dimensional Rasheed spacetime is a solution in Kaluza-Klein theory, which is obtained from a five-dimensional spacetime by dimensional reduction. One could argue that in this framework, the Hamilton-Jacobi equation for the four-dimensional problem, does not correctly describe the particle motion. Instead the equation of motion in the five-dimensional metric has to be considered \cite{Kovacs:1984qx}--\cite{Kerner:2000iz}. Here the dilatonic force yields new terms in the equations of motion.

In the following we will prove that in the above special cases, the particle motion is correctly described by the Hamilton-Jacobi equation for the four-dimensional problem.

In case 1, all new terms entering the equations of motion vanish since $\delta=0$ and $q=0$, compare \cite{Kovacs:1984qx}--\cite{Liu:1997fg}. In case 2, the dilaton field vanishes everywhere, because $J=0$. In case three, the dilaton field vanishes in the equatorial plane since $\theta=\frac{\pi}{2}$. However, derivatives of the dilaton field enter the equations of motion in the five-dimensional spacetime \cite{Kovacs:1984qx}--\cite{Liu:1997fg}. Therefore it has to be checked, if there are effects due to the dilaton in the equatorial plane.

The four-dimensional black hole spacetime $\dd s^2_{(4)}$ of Rasheed (see equation \eqref{eqn:ds4}) can be obtained from the five-dimensional metric
\begin{equation}
 \dd s^2_{(5)} = \frac{B}{A} \left( \dd x^5 + 2A_\mu\dd x^\mu \right) + \sqrt{\frac{A}{B}} \dd s^2_{(4)} = \tilde{g}_{\mu\nu} \dd x^\mu \dd x^\nu \, .
\end{equation}
The motion of a test particle in this metric is described by the Hamilton-Jacobi equation
\begin{equation}
 \half \tilde{g}^{\mu\nu} \tilde{p}_\mu \tilde{p}_\nu + \frac{\partial S_5}{\partial \lambda} =0 
\label{eqn:HJD5}
\end{equation}
with $\tilde{p}_\mu = \frac{\partial S_5}{\partial x^\mu}$. To solve equation \eqref{eqn:HJD5} we propose the ansatz
\begin{equation}
 S_5= \half\delta\lambda - Et + L\phi + K w + S_\theta(\theta) + S_r(r)
\end{equation}
where $K$ is a constant of motion corresponding to the momentum in the direction if the fifth dimension $x^5=w$. With this ansatz equation \eqref{eqn:HJD5} becomes
\begin{equation}
   \delta + \tilde{g}^{tt}E^2-2 \tilde{g}^{t\phi}EL+ \tilde{g}^{\phi\phi}L^2 +  \tilde{g}^{ww}K^2-2 \tilde{g}^{tw}EK + 2 \tilde{g}^{\phi w}K+  \tilde{g}^{\theta\theta } \left(  \frac{\partial S_\theta(\theta)}{\partial \theta} \right)^2 + \tilde{g}^{rr} \left(  \frac{\partial S_r(r)}{\partial r} \right)^2 = 0
\label{eqn:HJD5-2}
\end{equation}
Inserting the metric functions of the five-dimensional metric into equation \eqref{eqn:HJD5-2} results in a long and complicated expression, which we will not display here. Setting then $\theta=\frac{\pi}{2}$ simplifies the equation to
\begin{align}
 &\delta r^2 + \frac{(r^4-4 J^2) E^2}{(r-M)^2} + \frac{4JEL}{r-M} - L^2 +(r-M)^2 \left(  \frac{\partial S_r(r)}{\partial r} \right)^2 - \frac{(2Mr^5-3M^2r^4 -r^6 + 32J^2M^2 - 32J^2Mr + 8J^2r^2)K^2}{r^2(r-M)^2} \nonumber\\ 
 &+ \frac{ ((4r-8M)J^2+Mr^4) 2\sqrt{2}EK}{r(r-M)^2} + \frac{ 4J(2M-r) \sqrt{2}LK}{r(r-M)} = 0 \, .
\label{eqn:HJD5-equa}
\end{align}
It can be shown that equation \eqref{eqn:HJD5-equa} is the same as the Hamilton-Jacobi equation \eqref{eqn:HJD4-equa} for a charged particle in the equatorial plane of the four-dimensional spacetime if we make the identifications
\begin{equation}
  q=2K \quad \text{and} \quad \delta=m^2=1+K^2
\end{equation}
which result in the relation
\begin{equation}
 \frac{q}{m} = \frac{2K}{\sqrt{1+K^2}} \ .
\end{equation}
The same relation without the factor 2 was found in \cite{Liu:1997fg}, \cite{Kerner:2000iz}. Here the factor 2 is due to the fact that there is an additional factor 2 in front of the vector field $A_\mu$ in the five-dimensional Rasheed metric, which is not present in the metric in \cite{Liu:1997fg}, \cite{Kerner:2000iz}. Therefore in the equatorial plane of the extreme Rasheed metric, the Hamilton-Jacobi equation from the four-dimensional particle motion agrees with the Hamilton-Jacobi equation from the five-dimensional problem and correctly describes the motion of test particles.
\\

In this article we are interested in the motion of charged particles around an extremal rotating dyonic black hole in Kaluza-Klein theory; thus, we will consider the third case and set $\theta = \frac{\pi}{2}$ from now on. The equations of motion in the equatorial plane can be derived from the action $S$ \eqref{eqn:action}
\begin{align}
 \left( \frac{\dd r}{\dd\gamma} \right)^2 &= R(r) = \sum_{i=0}^6 a_i r^i  \, , \label{eqn:r-equation}\\
 \left( \frac{\dd\phi}{\dd\gamma}\right) &= Lr + \frac{1}{r-1}\left[ Jr\left( \sqrt{2}q-2E \right) - 2\sqrt{2}Jq \right]  \, ,  \label{eqn:phi-equation}\\
 \left( \frac{\dd t}{\dd\gamma}  \right) &=  \frac{1}{(r-1)^2}\left[ 2JLr(r-1) + Er(r^4-4J^2) + \sqrt{2}q \left( \half r^4 + 2J^2(r-2) \right) \right] \, .  \label{eqn:t-equation}
\end{align}
To simplify the equations of motion we used dimensionless quantities equivalent to setting $M=1$
\begin{equation}
 r\rightarrow Mr \, , \ t\rightarrow Mt \, , \ \lambda\rightarrow M\lambda\ \, , \ J\rightarrow M^2J \, , \ L\rightarrow ML
\end{equation}
and we also applied the Mino-time \cite{Mino:2003yg} $\gamma$ with $\dd\lambda = r^3 \dd \gamma$. The function $R(r)$ in equation \eqref{eqn:r-equation} is a polynomial of order $6$ with the coefficients
\begin{align}
 a_6 &= E^2-\delta\, ,\\
 a_5 &= \sqrt{2}qE+2\delta\, ,\\
 a_4 &= \half q^2-\delta-L^2\, ,\\
 a_3 &= 2L\left(2EJ-\sqrt{2}qJ+L\right)\, ,\\
 a_2 &= -J^2\left(2E-\sqrt{2}q\right)^2-2LJ\left(2E-3\sqrt{2}q\right)-L^2\, ,\\
 a_1 &= -4Jq\left(2\sqrt{2}EJ+\sqrt{2}L-2qJ\right)\, ,\\
 a_0 &= -q^2J^2\, .
\end{align}
In general the equations of motion are of hyperelliptic type. However, in the special case of uncharged particles $q=0$ the equations of motion reduce to elliptic type (note that $\dd\lambda = r \dd \gamma$ in this case)
\begin{align}
 \left( \frac{\dd r}{\dd\gamma} \right)^2 &= R(r) = \sum_{i=0}^4 a_i r^i  \, , \label{eqn:r-equation2}\\
 \left( \frac{\dd\phi}{\dd\gamma}\right) &= L - \frac{2JE }{r-1}  \, ,  \label{eqn:phi-equation2}\\
 \left( \frac{\dd t}{\dd\gamma}  \right) &=  \frac{2JL(r-1) + E(r^4-4J^2)}{(r-1)^2} \, . \label{eqn:t-equation2}
\end{align}
where
\begin{align}
 a_4 &= E^2-\delta\, ,\\
 a_3 &= 2\delta\, ,\\
 a_2 &= -\delta-L^2\, ,\\
 a_1 &= 2L(2EJ+L)\, ,\\
 a_0 &= -(2EJ+L)^2-2ELJ\, .
\end{align}
Both in the hyperelliptic and the elliptic case the equations of motion can be solved analytically as shown in section \ref{sec:solutions}.\\

\section{Classification of the orbits}

In this section we will analyze the motion of charged particles in the spacetime of the extremal black hole found by Rasheed in Kaluza-Klein theory.

\subsection{The azimuthal motion}
\label{sec:azimuth}
The $\phi$-equation \eqref{eqn:phi-equation} determines the azimuthal motion. $\dd\phi / \dd\gamma$ vanishes if
\begin{equation}
 E_{\rm turn} = \frac{Jq\sqrt{2}(r-2)+Lr(r-1)}{2Jr}
\label{eqn:eturn}
\end{equation}
or
\begin{equation}
 r^{\rm turn}_{1,2}=\frac{1}{2L} \left[ L-\sqrt{2}qJ+2JE  \pm \sqrt{ \left(  L-\sqrt{2}qJ+2JE \right)^2 + 2\sqrt{2}JL  }  \right] \, .
\end{equation}
At these radii the test particles change their direction, which leads to interesting orbits, see section \ref{sec:orbits}. A similar effect is usually observed at the ergoregion of a black hole. However, in this spacetime an ergoregion does not exist. The radii at which the angular direction changes were called \emph{turnaround boundaries} in \cite{Diemer:2013fza}.

\subsection{The radial motion}

The radial equation of motion determines the type of the orbits, because its zeros are the turning points of the orbit. Particle motion is allowed for $R(r)\geq 0$ only. Here we are interested in the real positive zeros. To study the possible orbits, parametric diagrams and effective potentials can be constructed from the polynomial $R(r)$. First we will give a list of the possible orbit types:
\begin{enumerate}
 \item Bound orbits (BO) with $r\in [r_1, r_2]$ exist for $r_{1,2} > r_H$ and also for $r_{1,2}  < r_H$ hidden behind the event horizon.
 \item Many-world bound orbits (MBO) with $r\in [r_1, r_2]$ and $r_1 < r_H < r_2$, where the particle crosses the horizon several times. Each time the horizon is crossed twice the particle enters another universe.
 \item Escape orbits (EO) with  $r\in [r_1, \infty)$ and $r_1 > r_H$, where the test particle (or light) escapes the gravity of the black hole.
 \item Two-world escape orbits (TWEO) with  $r\in [r_1, \infty)$ and $r_1 < r_H$, where the particle crosses the horizon twice and enters another universe.
\end{enumerate}

\begin{figure}[h!]
  \includegraphics[width=0.4\linewidth]{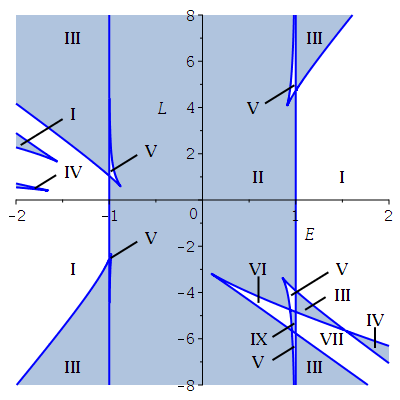}
  \caption{$E$-$L$-Parameterplot for charged particles with $\delta=1, J=0.4, q=0.8$. The blue curves separate regions with a different number of zeros of $R$, and therefore different types of orbits. The roman numbers refer to these orbit types and are explained in table \ref{tab:orbits-del1}.}
\label{pic:parameterplot-del1}
\end{figure}

The number of zeros of $R(r)$ for a set of parameters is closely related to the possible orbit types. If double zeros occur, i.e. $R(r)=0$ and $\frac{\dd R}{\dd r}=0$, the number of zeros changes. We plot these two conditions as a parameterplot of $E$ and $L$ to see nine regions with different numbers of real positive zeros. An example of a parameterplot is shown in figure \ref{pic:parameterplot-del1}. 
There is one real positive zero in region I, two zeros in region II, three zeros in the regions III and IV,  four zeros in the regions V and VI, five zeros in the regions VII and VII, and six zeros in region IX. Two regions may have the same number of positive real zeros; however, there are different types of possible orbits. 

The orbit types in each region can be determined with the help of effective potentials. We define an effective potential consisting of two parts $V^\pm$ by
\begin{equation}
 \left( \frac{\dd r}{\dd \gamma} \right)^2 = R(r)= \left(r^6-4J^2r^2 \right)  \left(E-V^+ \right)  \left(E-V^- \right)
\end{equation}
and therefore
\begin{equation}
 V^\pm = -\frac{q}{\sqrt{2}r} + \frac{r-1}{r \left( 4J^2-r^4 \right)}\left[ 2J\left(rL+\sqrt{2}qJ\right) \pm r^2\sqrt{ \left(rL+\sqrt{2}qJ\right)^2 -\delta\left( 4J^2-r^4 \right)  } \right] \, .
 \label{eqn:potential}
\end{equation}
In the case of uncharged photons $\delta=q=0$ the effective potentials simplify to
\begin{equation}
 V^\pm = \frac{L(r-1)}{2J\mp r^2}\, .
\end{equation}
At the horizon $r_H=1$ the two effective potentials meet at $E=-\frac{q}{\sqrt{2}}$. Here $R(r=r_H, E=-\frac{q}{\sqrt{2}})=0$ and a circular orbit exists directly on the horizon. Furthermore, the two potentials meet at a point behind the horizon, which can be calculated by
\begin{equation}
 \left(rL+\sqrt{2}qJ\right)^2 = \delta\left( 4J^2-r^4 \right) \, .
\end{equation}
For $r\rightarrow 0$ both potentials diverge. Additionally for $r\rightarrow \sqrt{2|J|}$, the potential $V_+$ diverges if $JL<0$ and $V_-$ diverges if $JL>0$. However, the latter divergency has no effect on the orbits, since the equations of motion remain smooth and therefore we will not show this divergency in the plots of the effective potentials . At $r\rightarrow \infty$ the potentials $V^ \pm$ aproach $\mp \sqrt{\delta}$ asymptotically. We find the following symmetries
\begin{align}
 V^\pm (-J, -L) &=  V^\pm (J, L) \, , \nonumber \\
 V^\pm (-q, -L) &=  -V^\mp (q, L) \, ,\nonumber \\
 V^\pm (-q, -J) &=  -V^\mp (q, J) \, . 
\label{eqn:symmetry}
\end{align}
In the Schwarzschild spacetime the positive root of the effective potential is always $V_+\geq0$ and the negative root $V_-\leq0$. There is no classical interpretation for particles with negative energies; however, in quantum field theory they can be associated with antiparticles \cite{Deruelle:1974zy}. Note that in contrast to the Schwarzschild spacetime, here the positive root $V_+$ can be negative, which can allow particles with negative energies. In \cite{Christodoulou:1972kt, Denardo:1973} the region for particles with negative energies was described as a ``generalized ergosphere'', where energy may be extracted. Also the negative root $V_-$ can be positive in this spacetime. To find a bound which separates particles and antiparticles, one can consider equation \eqref{eqn:t-equation} which describes the coordinate time $t$. Time should run forward for particles and backwards for antiparticles. Solving the $t$-equation for $E$ will give a lower energy bound for particles, which can be negative in the ``generalized ergosphere''.

\begin{figure}[hp]
	\subfigure[~$\delta=1, q=0.8, J=0.4, L=-3.75$]{
		\includegraphics[width=0.4\linewidth]{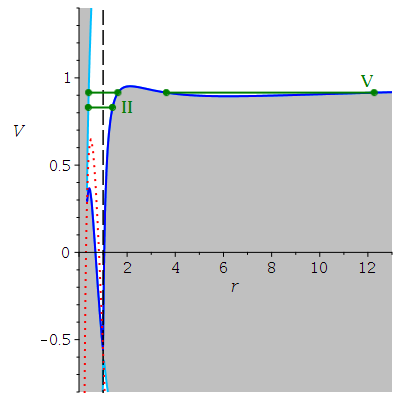}
	}
	\subfigure[~Closeup of figure (a)]{
		\includegraphics[width=0.4\linewidth]{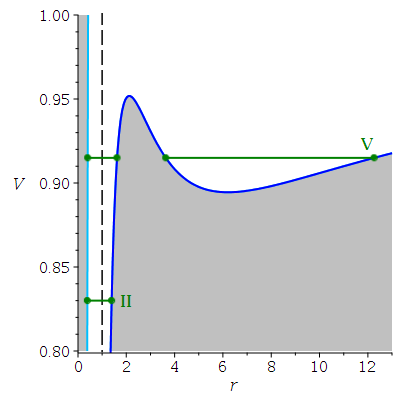}
	}
	\subfigure[~$\delta=1, q=0.95, J=0.3, L=-4.4$]{
		\includegraphics[width=0.4\linewidth]{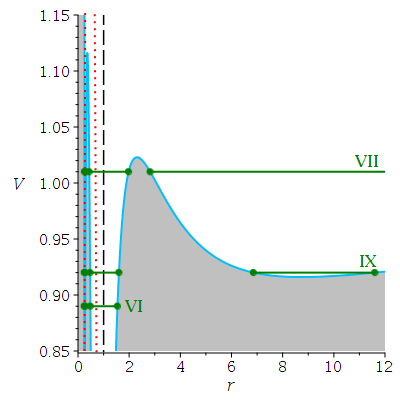}
	}
	\subfigure[~Closeup of figure (b)]{
		\includegraphics[width=0.4\linewidth]{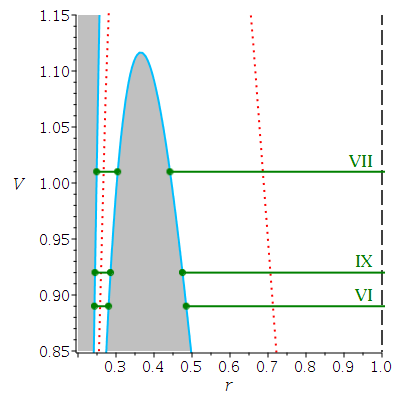}
	}
	\subfigure[~$\delta=1, q=0.5, J=0.4, L=-27$]{
		\includegraphics[width=0.4\linewidth]{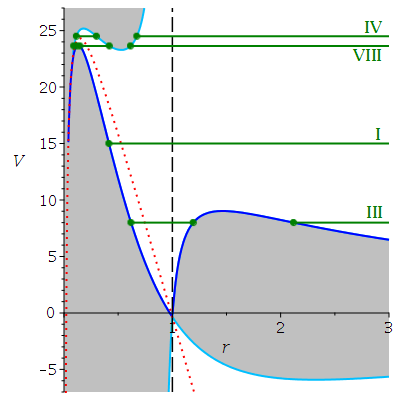}
	}
	\subfigure[~Closeup of figure (e)]{
		\includegraphics[width=0.4\linewidth]{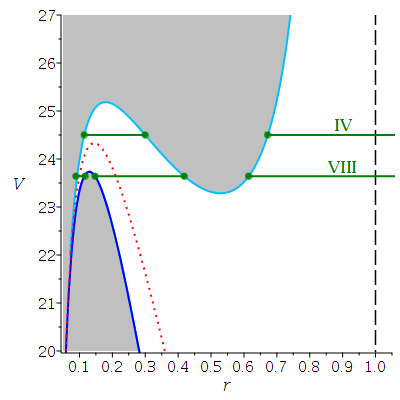}
	}
  \caption{Plot of the effective potentials $V_+$ (light blue curves) and $V_-$ (dark blue curves) along with some example energies (green lines) for different orbits. The green circles represent the turning points. The roman numbers refer to the orbit types and are explained in table \ref{tab:orbits-del1}. Particle motion is only possible in the white area, in the grey area the motion is forbidden due to $R<0$. The red dotted curves show the energy $E_{\rm turn}$ (equation \eqref{eqn:eturn}), where a turnaround boundary occurs. The vertical black dashed lines indicate the position of the horizon.}
\label{pic:potential}
\end{figure}

Some examples of the potentials are depicted in figure \ref{pic:potential}. We also plot the energy $E_{\rm turn}$ (equation \eqref{eqn:eturn}) where the test particles changes angular direction.\\

Taking all the information from the parameterplots and the effective potentials into account, we determine the possible orbits for charged particles in the spacetime of an extremal black hole in Kaluza-Klein theory. An overview of the regions and the corresponding orbit types can be found in table \ref{tab:orbits-del1} and the following list.
\begin{enumerate}
 \item Region I: The polynomial $R$ has one zero $r_1 < r_H$ and the corresponding orbit is a TWEO.
 \item Region II: $R$ has two zeros $r_1 < r_H$ and $r_2 > r_H$. The corresponding orbit is a MBO.
 \item Region III: $R$ has three zeros $r_1 < r_H$ and $r_2 ,r_3 > r_H$. A MBO and a EO exist.
 \item Region IV:  $R$ has three zeros $r_1, r_2 < r_H$ and $r_3 > r_H$. A BO hidden behind the horizon and a TWEO exist.
 \item Region V: $R$ has four zeros  $r_1 < r_H$ and $r_2 , r_3 , r_4  > r_H $. Here a MBO and a BO exist.
 \item Region VI: $R$ has four zeros  $r_1 , r_2 , r_3 < r_H$ and $r_4 > r_H$. A BO hidden behind the horizon and a MBO exist.
 \item Region VII: $R$ has five zeros  $r_1 , r_2 , r_3 < r_H$ and $r_4, r_5 > r_H$.  A BO hidden behind the horizon, a MBO and a EO exist.
 \item Region VIII:  $R$ has five zeros  $r_1 , r_2 , r_3, r_4, r_5  < r_H$.  Two BOs hidden behind the horizon and a TWEO exist.
 \item Region IX: $R$ has six zeros  $r_1 , r_2 , r_3 < r_H$ and $r_4, r_5, r_6 > r_H$.  A BO hidden behind the horizon, a MBO and a second BO outside the black hole exist.
\end{enumerate}
In the special case of massless particles with $\delta=q=0$, only the regions I, II, III and IV exist. For massless particles in region II, $R$ possesses a double zero at $r_H=1$ and the energy is always $E=0$, here light moves on a circular orbit directly on the horizon.\\

\begin{table}[h!]
  \includegraphics[width=0.6\linewidth]{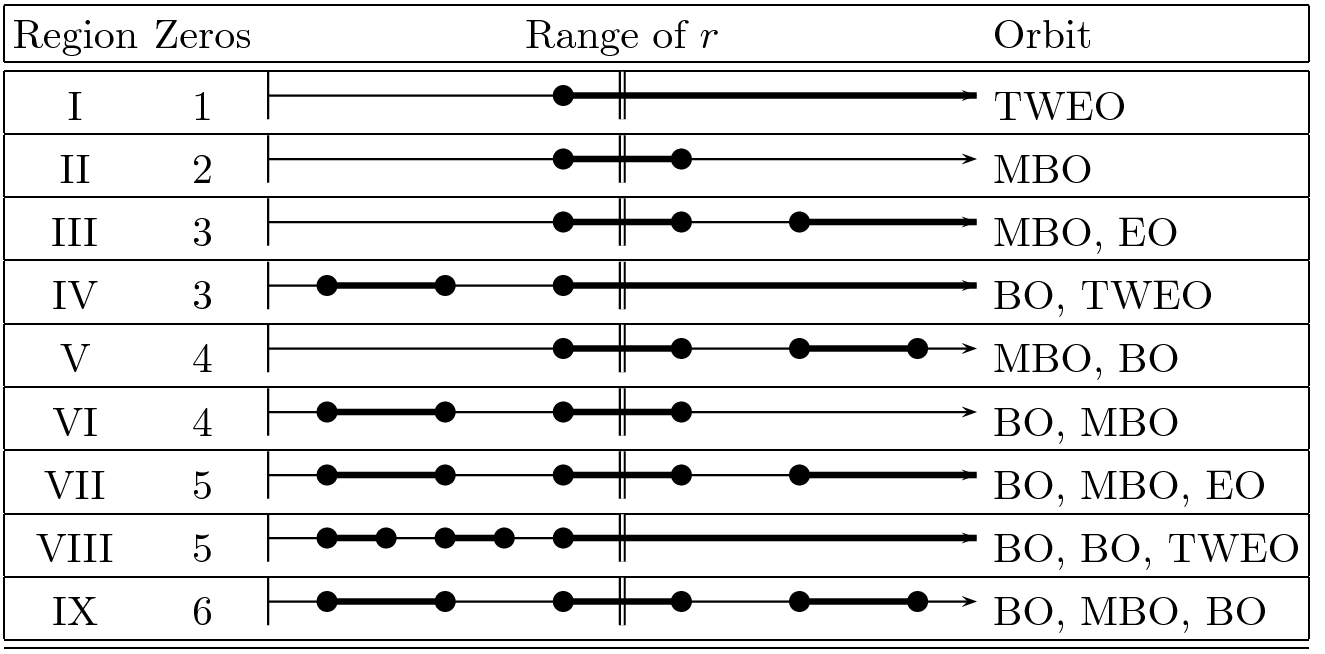}
  \caption{Types of orbits for charged particles in the spacetime of an extremal black hole in Kaluza-Klein theory. The thick lines show the range of the orbits and the thick dots mark the turning points. The single vertical line corresponds to the singularity and the vertical double lines indicate the position of the horizon.}
  \label{tab:orbits-del1}
\end{table}

Let us now investigate under which conditions an orbit terminates at the singularity $r_S$. At the singularity, the polynomial $R$ has the value
\begin{equation}
 R(r_S) = -8 J^2 q^2\, .
\end{equation}
Therefore in the case $J\neq 0$ and $q\neq0$ particles can never reach the singularity. If $J=0$ or $q=0$ we have $R(r_S)=0$, but then $r_S=0$ is a local maximum of $R$, so that test particles can never reach that point.
\begin{enumerate}
 \item If $J=0$ then $R'(r=0)=0$ and $R''(r=0)=-L^2$.
 \item If $q=0$ then $R'(r=0)=0$ and $R''(r=0)=-2(2EJ+L)^2$.
\end{enumerate}
If additionally $L=0$ in the first case or $J=L=0$ in the second case then $r_S=0$ is an inflection point. The effective potentials show that the singularity is guarded by a potential barrier which makes it impossible to find orbits that reach $r_S=0$. Only in the special case $J=L=q=\delta=0$ particles can reach the singularity. Here in fact all orbits end in the singularity.\\

\subsection{Static orbits}

In \cite{Collodel:2017end} static orbits in axisymmetric rotating spacetimes were discovered. Under certain conditions there is a ring of points in the equatorial plane where particles remain at rest. We will show that these orbits also exist in the Rasheed spacetime.

A particle is at rest at a certain point if 
\begin{align}
 \frac{\dd r}{\dd \gamma}&=0 \, , \nonumber\\
 \frac{\dd \phi}{\dd \gamma}&=0 \, .
\end{align}
From this we can calculate the radius and the energy of the particle at rest
\begin{align}
 r_{\rm rest} &= \frac{J}{L}\left(-\sqrt{2}q\pm 2\sqrt{\delta}\right)\, ,\nonumber\\
 E_{\rm rest} &= \frac{ \pm \sqrt{\delta}\left(2J\sqrt{2}q + 2L\right) + L\sqrt{2}q - 4J\delta}{2J\left(\sqrt{2}q\pm 2\sqrt{\delta}\right)} \, .
\end{align}
At $r_{\rm rest}$ the effective potentials (equation \eqref{eqn:potential}) intersect with the turnaround boundary (equation \eqref{eqn:eturn}) and the particle is at rest at the pericenter or apocenter of its orbit. If the turnaround boundary intersects with the local minimum of one of the effective potentials, i.e. if
\begin{align}
  \frac{\dd r}{\dd \gamma}&=0 \, , \nonumber\\
  \frac{\dd^2 r}{\dd \gamma^2}&=0 \, , \nonumber\\
  \frac{\dd \phi}{\dd \gamma}&=0 \, ,
\end{align}
the particle remains at rest at all times, this is called the \emph{static ring}. We set $\delta=1$, because then there will be a local minimum in one of the effective potentials for $r>r_H$. The above equations then yield
\begin{align}
 q_{\rm st}&=\mp\sqrt{2} \, , \nonumber \\
 E_{\rm st}&=\pm 1 \, , \nonumber \\
 r_{\rm st}&=\pm 4\frac{J}{L}\, . \\
\end{align}
Additionally the particle remains at rest at the points where the two effective potentials $V^\pm$ meet, including the horizon $r_H=1$.

\begin{figure}[h]
	\subfigure[~$\delta=1, q=-0.5, J=0.4, L=-0.75$]{
		\includegraphics[width=0.31\linewidth]{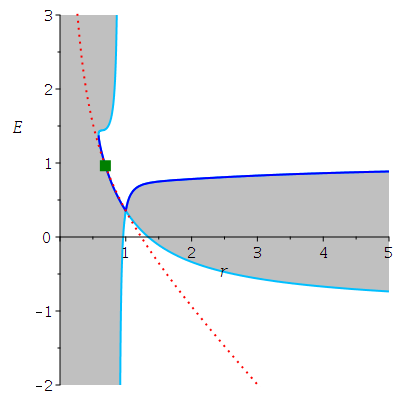}
	}
	\subfigure[~$\delta=1, q=-0.5, J=0.4, L=0.25$]{
		\includegraphics[width=0.31\linewidth]{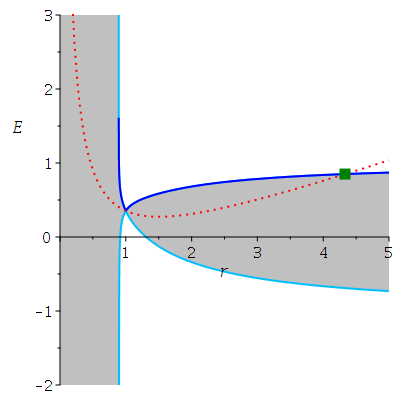}
	}
	\subfigure[~$\delta=1, q=-\sqrt{2}, J=0.49, L=1.5$]{
		\includegraphics[width=0.31\linewidth]{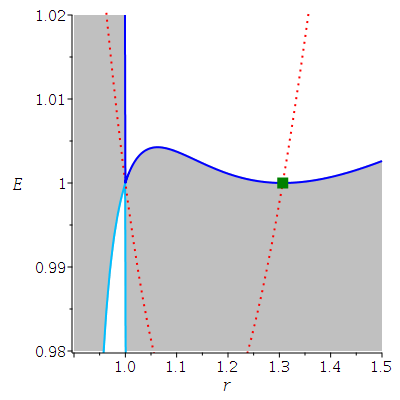}
	}
  \caption{Plots of the effective potentials  $V_+$ (light blue curves), $V_-$ (dark blue curves) and the energy $E_{\rm turn}$ (red dotted curve) of the turnaround boundary. In the grey area motion is forbidden since $R<0$. At the green squares $V^\pm$ and $E_{\rm turn}$ intersect, so that $\frac{\dd r}{\dd \gamma}=\frac{\dd \phi}{\dd \gamma}=0$ and the particle is at rest. In subfigure (a) and (b) the particle is at rest at the pericenter or apocenter, respectively. Subfigure (c) shows the static ring, here the particle remains at rest at all times.}
\label{pic:static_pot}
\end{figure}

Figure \ref{pic:static_pot} shows some examples of static orbits. There the effective potentials and the turnaround boundary are depicted as well as the position ($r$ and $E$) of a particle at rest. In figure \ref{pic:static_pot}(a) and \ref{pic:static_pot}(b) the particle is at rest at the pericenter or apocenter, respectively. Figure \ref{pic:static_pot}(c) shows the static ring, here the particle remains at rest at all times.

\subsection{Photon sphere}

The photon sphere is a radius around the black hole, where light moves on unstable circular orbits (see e.g. \cite{Claudel:2000yi}). It marks the region between light rays escaping the black hole and light rays falling beyond the event horizon. Via a coordinate transformation \cite{Grenzebach:2014fha}-\cite{Grenzebach:Springer} the shadow of a black hole can be obtained from the photon sphere.

We consider the $r$-equation for $\delta=0$ and $q=0$, equation \eqref{eqn:r-equation2}, and apply the conditions for unstable circular orbits
\begin{align}
 R&=0 \, , \nonumber\\
 \frac{\dd R}{\dd r}&=0 \, .
\end{align}
Solving the above equations yields
\begin{align}
 \frac{L}{E}&=\pm 2 r_{\rm ph} \, ,\\
 \frac{J}{E}&=\pm \left(\half  r_{\rm ph}^2-r_{\rm ph}\right) \, .
\end{align}
Then there are four solutions for the radius $r_{\rm ph}$ of the photon sphere which are valid for different ranges of the angular momentum of the black hole $J$ and the angular momentum of the photon $L$. There is always a photon sphere outside the event horizon $r_{\rm ph}\geq r_H$ and for certain ranges of $J$ a second $r_{\rm ph}$ exists hidden behind the horizon. The solutions for the photon sphere are:
\begin{itemize}
 \item $L>0$ and $-\half < J$: $r_{\rm ph}=1+\sqrt{1+2J}\geq r_H$
 \item $L>0$ and $-\half < J <0$: $r_{\rm ph}=1-\sqrt{1+2J} \leq r_H$
 \item $L<0$ and $ J < \half$:  $r_{\rm ph}=1+\sqrt{1-2J}\geq r_H$
 \item $L<0$ and  $ 0 < J < \half$:  $r_{\rm ph}=1-\sqrt{1-2J}\leq r_H$
\end{itemize}

\subsection{Causality and time flow}

Note that this spacetime allows closed timelike curves (CTCs). For an extensive discussion on CTCs see \cite{Gibbons:1999uv}. CTCs occur for $g_{\phi\phi}<0$. In this spacetime the metric component is
\begin{equation}
 g_{\phi\phi} = \frac{\left(r^4-4J^2\right) \sin^2\theta}{\sqrt{r^4-4J^2\cos^2\theta}}
\end{equation}
and therefore, CTC are present in the region $r_S< r <r_{CTC}$ with $r_{CTC}=\sqrt{2|J|}$. Since $r_{CTC}< r_H$ for $J< \frac{M^2}{2}$, the CTCs are hidden behind the degenerate horizon.\\ 

As in \cite{Gibbons:1999uv} we can consider an effective potential from the $t$-equation by solving \eqref{eqn:t-equation} for $E$
\begin{equation}
 V_{\rm time}= -\frac{q}{\sqrt{2}r} + \frac{r-1}{r \left( 4J^2-r^4 \right)}\left[ 2J\left(rL+\sqrt{2}qJ\right) \right] \, .
\end{equation}
$V_{\rm time}$ is equal to the first part (i.e. without the root) of the effective potentials $V^\pm$ from the $r$-equation. $V_{\rm time}$ diverges for $r\rightarrow r_{CTC}=\sqrt{2|J|}$. From the effective potential $V_{\rm time}$ we get information on the time flow of the coordinate time $t$ with respect to the affine parameter $\gamma$ (which is related to the eigentime). There are regions where  $\left( \frac{\dd t}{\dd\gamma}  \right) >0$ and regions where $\left( \frac{\dd t}{\dd\gamma}  \right) <0$. Time runs forward for particles and backwards for antiparticles.

\begin{figure}[h!]
	\subfigure[~$\delta=1, q=0.8, J=0.4, L=-3.75$]{
		\includegraphics[width=0.4\linewidth]{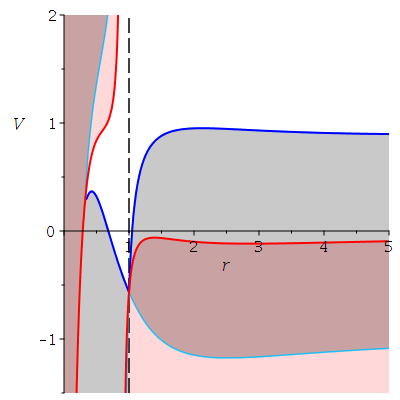}
	}
	\subfigure[~$\delta=1, q=0.5, J=0.4, L=-27$]{
		\includegraphics[width=0.4\linewidth]{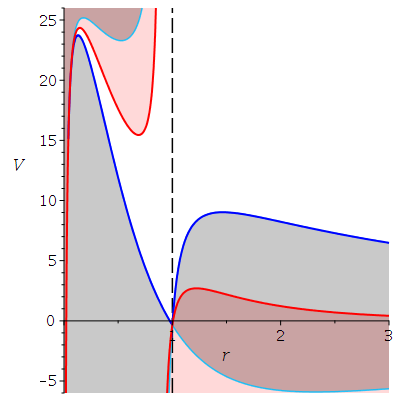}
	}
  \caption{Effective potentials  $V_+$ (light blue curves) and $V_-$ (dark blue curves) from the $r$-equation and from the $t$-equation (red). The grey areas are forbidden since $R<0$. In the red areas the coordinate time runs backwards $\left( \frac{\dd t}{\dd\gamma}  \right) <0$. The vertical dashed line indicates the position of the horizon.}
\label{pic:timepotential}
\end{figure}

In figure \ref{pic:timepotential} we plot the effective potentials from the $r$-equation together with the potential from the $t$-equation. We also see here that there are particles with  $\left( \frac{\dd t}{\dd\gamma}  \right) >0$, but with negative energies. In \cite{Christodoulou:1972kt, Denardo:1973} this region was called ``generalized ergosphere''. 

For most orbits, the time flow stays either positive or negative. However, for certain sets of parameters, particles can cross from a region with $\left( \frac{\dd t}{\dd\gamma}  \right) >0$ into a region with  $\left( \frac{\dd t}{\dd\gamma}  \right) <0$, so that the time flow is reversed at at some point in these orbits. This occurs behind the horizon only.

\section{Solution of the equations of motion}
\label{sec:solutions}
In this section we will present the analytical solution of the equations of motion for charged and uncharged particles in the equatorial plane of an extremal black hole in Kaluza-Klein theory. The solutions of the $r$-equation and the $\phi$-equation can then be used to plot the orbits. For charged particles the equations of motion \eqref{eqn:r-equation}--\eqref{eqn:t-equation} are of hyperelliptic type. However, in the special case of uncharged particles $q=0$ the equations of motion reduce to elliptic type, see equations \eqref{eqn:r-equation2}--\eqref{eqn:t-equation2}.

\subsection{The $r$-equation}
\label{sec:r-solution}
\subsubsection{Hyperelliptic case}
For charged particles the right side of the $r$-equation \eqref{eqn:r-equation} is a polynomial of order 6
\begin{equation}
  \left( \frac{\dd r}{\dd\gamma} \right)^2 = R(r) = \sum_{i=0}^6 a_i r^i \, .
\end{equation}
The order of the polynomial can be reduced by substituting $r=\pm\frac{1}{x} + r_0$ with $R(r_0)=0$
\begin{equation}
  \left( x \frac{\dd x}{\dd\gamma} \right)^2 = \sum_{i=0}^5 b_i x^i \quad \text{with} \quad b_i=\left. \frac{(\pm 1)^i}{(6-i)!}\frac{\dd^{6-i}R}{\dd r^{6-i}}\right| _{r=r_0} \, .
\end{equation}
A second substitution $x=\frac{4}{b_5}y$ transforms the polynomial into the canonical form
\begin{equation}
  \left( y \frac{\dd y}{\dd\gamma} \right)^2 = \sum_{i=0}^5 \lambda_i y^i = P_5(y)\, 
\end{equation}
where
\begin{equation}
 \lambda_5 = 4 \, , \ \lambda_4=b_4 \, , \ \lambda_3 = \frac{b_3 b_5}{4} \, , \ \lambda_2 = \frac{b_2 b_5^2}{16} \, , \ \lambda_1 = \frac{b_1 b_5^3}{64} \, , \ \lambda_0 = \frac{b_0 b_5^4}{256} \, .
\end{equation}
A separation of variables leads to a hyperelliptic integral of genus 2
\begin{equation}
 \gamma - \gamma_{\rm in} = \int_{y_{\rm in}}^y \frac{y\dd y}{\sqrt{P_5(y)}}
\end{equation}
where $y_{\rm in}=\frac{\pm b_5}{4(r-r_{\rm in})}$ and $r_{\rm in}$ is the initial value for $r$. We are looking for a solution $y(\gamma)$, so basically we have to solve a special case of the Jacobi inversion problem, see \cite{Hackmann:2008tu}, \cite{Hackmann:2008zz}, \cite{Enolski:2010if}, \cite{Enolski:2011id} for a detailed discussion. The solution is
\begin{equation}
 y(\gamma) = \left. \frac{\sigma_1\left( \vec{\gamma}_\infty \right)}{\sigma_2\left( \vec{\gamma}_\infty \right)} \right| _{\sigma\left( \vec{\gamma}_\infty \right)=0}
\end{equation}
where
\begin{equation}
  \vec{\gamma}_\infty = \left(
    \begin{array}{c}
      \gamma_1\\
      \gamma-\gamma_{\rm in}'
    \end{array}
  \right) 
  \quad \text{and} \quad
  \gamma_{\rm in}'=\gamma_{\rm in} + \int_{y_{\rm in}}^\infty \frac{y\dd y}{\sqrt{P_5(y)}} \, .
\end{equation}
If the initial value $r_{\rm in}$ is chosen at one of the turning points in the orbit, i.e. the zeros of $R(r)$, then  $\gamma_{\rm in}'$ can be expressed in terms of the periods. The functions $\sigma_1$ and $\sigma_2$ are derivatives of the Kleinian $\sigma$-function $\sigma_i=\frac{\partial \sigma (\vec{z})}{\partial z_i}$. The constant $\gamma_1$ can be determined by the condition $\sigma\left( \vec{\gamma}_\infty \right)=0$. By resubstitution we get the solution of the $r$-equation \eqref{eqn:r-equation}
\begin{equation}
 r(\gamma)=\pm \frac{b_5}{4 } \left. \frac{\sigma_2\left( \vec{\gamma}_\infty \right)}{\sigma_1\left( \vec{\gamma}_\infty \right)} \right| _{\sigma\left( \vec{\gamma}_\infty \right)=0} + r_0\, .
\end{equation}

\subsubsection{Elliptic case}
For neutral particles the $r$-equation simplifies so that the right side of equation \eqref{eqn:r-equation2} is a polynomial of order 4
\begin{equation}
  \left( \frac{\dd r}{\dd\gamma} \right)^2 = R(r) = \sum_{i=0}^4 a_i r^i
\end{equation}
which can be reduced to third order by the substitution $r=\pm\frac{1}{x} + r_0$ with $R(r_0)=0$
\begin{equation}
	\left(\frac{\dd x}{\dd \gamma}\right) ^2 = \sum _{i=1}^3 b_i x^i \, .
\end{equation} 
Substituting further $x=\frac{1}{b_3}\left( 4y-\frac{b_2}{3}\right)$ gives the standard Weierstra{\ss} form
\begin{equation}
	\left(\frac{\dd y}{\dd \gamma}\right)^2= 4 y^3-g_2 y -g_3 = P_3 (y)
	\label{eqn:weierstrass-form}
\end{equation}
with the coefficients
\begin{equation}
	g_2=\frac{b_2^2}{12} - \frac{b_1b_3}{4} \, , \qquad  g_3=\frac{b_1b_2b_3}{48} - \frac{b_0b_3^2}{16}-\frac{b_2^3}{216} \ .
\end{equation}
Equation \eqref{eqn:weierstrass-form} is solved by the Weierstra{\ss} $\wp$-function and therefore after resubstitution the solution of $r$-equation \eqref{eqn:r-equation2} is
\begin{equation}
	r(\gamma)=\pm \frac{b_3}{4 \wp\left(\gamma - \gamma'_{\rm in}; g_2, g_3\right) - \frac{b_2}{3}} +r_0
\end{equation}
with the initial values $\gamma'_{\rm in}=\gamma_{\rm in}+\int^\infty_{y_{\rm in}}{\frac{\dd y}{\sqrt{4y^3-g_2y-g_3}}}$ and $y_{\rm in}=\pm\frac{b_3}{4(r_{\rm in}-r_0)} + \frac{b_2}{12}$.

\subsection{The $\phi$-equation}
\label{sec:phi-solution}
\subsubsection{Hyperelliptic case}
Let us first solve the $\phi$-equation in case of charged particles. We use the substitution $r=\pm \frac{b_5}{4y}+r_0$ and knowing that $\dd\gamma=\frac{y\dd y}{\sqrt{P_5(y)}}$ we can write the $\phi$-equation \eqref{eqn:phi-equation} as
\begin{equation}
 \dd \phi = \left( C_0 + C_1 y + \frac{C_2}{y-p} \right) \frac{\dd y}{\sqrt{P_5(y)}}
 \label{eqn:phi-partial}
\end{equation}
with the pole
\begin{equation}
 p=\frac{b_5}{4(r_0-1)}
\end{equation}
and the constants
\begin{align}
 C_0 &= -\frac{b_5}{4(r_0-1)^2}\left[ L(r_0-1)^2 + J(2E+\sqrt{2}q)  \right] \\
 C_1 &= \frac{1}{1-r_0}\left[ 2EJr_0+Lr_0(1-r_0)+Jq\sqrt{2}(2-r_0) \right]\\
 C_2 &= -\frac{b_5^2J(2E+\sqrt{2}q)}{(r_0-1)^3}
\end{align}
which arise from a partial fraction decomposition. Integrating equation \eqref{eqn:phi-partial} yields
\begin{align}
 \phi-\phi_{\rm in} &= C_0 \int_{y_{\rm in}}^y \! \frac{\dd y}{\sqrt{P_5(y)}} +  C_1 \int_{y_{\rm in}}^y \! \frac{y \dd y}{\sqrt{P_5(y)}} + C_2 \int_{y_{\rm in}}^y \! \frac{1}{y-p}\frac{\dd y}{\sqrt{P_5(y)}} \nonumber \\
 &= C_0 (\gamma_1 -\gamma_{\rm in}'') +  C_1 (\gamma -\gamma_{\rm in}') + C_2 \int_{y_{\rm in}}^y \! \frac{1}{y-p}\frac{\dd y}{\sqrt{P_5(y)}}
\end{align}
where $\gamma_{\rm in}'=\gamma_{\rm in} + \int_{y_{\rm in}}^\infty \frac{y\dd y}{\sqrt{P_5(y)}}$ and  $\gamma_{\rm in}''=-\int_{y_{\rm in}}^\infty \frac{\dd y}{\sqrt{P_5(y)}}$ can be expressed in terms of the periods if $y_{\rm in}$ is chosen to be a zero of $P_5$. As before $\gamma_1$ is the constant to be determined by $\sigma\left( \vec{\gamma}_\infty \right)=0$. The solution of the remaining integral, which is a hyperelliptic integral of the third kind, was found in \cite{Enolski:2011id} and can be proven with the help of the Riemann vanishing theorem. In terms of the Kleinian $\sigma$-function the solution is
\begin{equation}
	\int_{y_{\rm in}}^y \frac{1}{y-p}\frac{\dd y}{\sqrt{ P_5(y)}} =  \frac{1}{\sqrt{ P_5(p)}} \left[ -2 \int_{y_{\rm in}}^y \dd\vec{u}^T  \int_{e_0}^p \dd\vec{r}
	+\ln \frac{\sigma\left(\int_{\infty}^y \dd\vec{u}- \int_{e_0}^{p} \dd \vec{u} - \vec{K}_\infty  \right)}{\sigma\left(\int_{\infty}^y \dd \vec{u}+ \int_{e_0}^{p} \dd \vec{u} - \vec{K}_\infty \right)}
	- \ln \frac{\sigma\left(\int_{\infty}^{y_{\rm in}} \dd \vec{u} - \int_{e_0}^{p} \dd \vec{u} - \vec{K}_\infty  \right)}{\sigma\left(\int_{\infty}^{y_{\rm in}} \dd\vec{u} + \int_{e_0}^{p} \dd \vec{u} - \vec{K}_\infty  \right)} \right] 
	\label{eqn:sol-thirdkind}
\end{equation}
where the components of the vectors $\dd \vec{u}$ and $\dd \vec{r}$ are
\begin{equation}
 \dd u_1 = \frac{\dd y}{\sqrt{P_5(y)}} \, , \ 
 \dd u_2 = \frac{y\dd y}{\sqrt{P_5(y)}} \, , \
 \dd r_1 = (\lambda_3 y+ 2\lambda_4 y^2 + 12 y^3) \frac{\dd y}{4\sqrt{P_5(y)}} \, , \
 \dd r_2 =  \frac{y^2\dd y}{\sqrt{P_5(y)}} \, . 
\end{equation}
$\vec{K}_\infty$ is the vector of Riemann constants and the basepoint $e_0$ is a zero of $P_5$. The integrals in equation \eqref{eqn:sol-thirdkind} can also be expressed as
\begin{equation}
  \int_\infty^y \dd \vec{u}=
   \left(
    \begin{array}{c}
      \gamma_1\\
      \gamma-\gamma_{\rm in}'
    \end{array}
  \right) 
 =\vec{\gamma}_\infty 
  \ , \quad
  \int_\infty^{y_{\rm in}} \dd \vec{u}=
   \left(
    \begin{array}{c}
      \gamma_{\rm in}''\\
      \gamma_{\rm in}-\gamma_{\rm in}'
    \end{array}
  \right) 
 =\vec{\gamma}_\infty^{\, \rm in}
 \quad \text{and} \quad
  \int_{y_{\rm in}}^y \dd \vec{u} = \vec{\gamma}_\infty - \vec{\gamma}_\infty^{\, \rm in}
   \ .
\end{equation}
We also define the constant vectors  and $\int_{e_0}^p \dd\vec{r}= \vec{c}_r$ and $\int_{e_0}^p \dd\vec{u}= \vec{c}_u$. Then we can write the solution of the $\phi$-equation \eqref{eqn:phi-equation} as
\begin{equation}
  \phi (\gamma) =\phi_{\rm in} + C_0 (\gamma_1 -\gamma_{\rm in}'') +  C_1 (\gamma -\gamma_{\rm in}') +  \frac{C_2}{\sqrt{ P_5(p)}} \left[ -2 \left( \vec{\gamma}_\infty - \vec{\gamma}_\infty^{\, \rm in} \right)^T  \vec{c}_r
	+\ln \frac{\sigma\left( \vec{\gamma}_\infty  - \vec{c}_u - \vec{K}_\infty  \right)}{\sigma\left( \vec{\gamma}_\infty  + \vec{c}_u - \vec{K}_\infty \right)}
	- \ln \frac{\sigma\left( \vec{\gamma}_\infty^{\, \rm in} - \vec{c}_u - \vec{K}_\infty  \right)}{\sigma\left( \vec{\gamma}_\infty^{\, \rm in} + \vec{c}_u - \vec{K}_\infty  \right)} \right] 
\end{equation}

\subsubsection{Elliptic case}

In the special case $q=0$, we use the substitutions from section \ref{sec:r-solution}, which transform the $r$-equation \eqref{eqn:r-equation2} into the Weierstra{\ss} form (equation \eqref{eqn:weierstrass-form}). Then we can write the $\phi$-equation \eqref{eqn:phi-equation2} as
\begin{equation}
 \dd\phi = C_0 \frac{\dd y}{\sqrt{P_3(y)}} + \frac{C_1}{1-p} \frac{\dd y}{\sqrt{P_3(y)}} 
\end{equation}
where
\begin{align}
 C_0 &=L-\frac{2EJ}{r_0-1} \, ,\\
 C_1 &=\frac{EJb_3}{2(r_0-1)^2}\, ,\\
 p &= \frac{b_2}{12}-\frac{b_3}{4(r_0-1)}\, .
\end{align}
Then we obtain
\begin{equation}
  \phi (\gamma) =\phi_{\rm in} + C_0 (\gamma -\gamma_{\rm in}) + \int_{y_{\rm in}}^y\!\frac{C_1}{1-p}  \frac{\dd y}{\sqrt{P_3(y)}} \, .
\end{equation}
The remaining integral is an elliptic integral of the third kind which can easily be integrated (see  e.g. \cite{Enolski:2011id}, \cite{Kagramanova:2010bk}, \cite{Grunau:2010gd}), so that the solution of the $\phi$-equation \eqref{eqn:phi-equation2} is in terms of the Weierstra{\ss} $\wp$, $\zeta$ and $\sigma$ functions
\begin{equation}
  \phi (\gamma) =\phi_{\rm in} + C_0 (\gamma -\gamma_{\rm in}) + \frac{C_1}{\wp'(v)}\left( 2\zeta(v)(\gamma -\gamma_{\rm in}') + \ln\frac{\sigma(\gamma -\gamma_{\rm in}'-v)}{\sigma(\gamma_{\rm in} -\gamma_{\rm in}'-v)} - \ln\frac{\sigma(\gamma -\gamma_{\rm in}'+v)}{\sigma(\gamma_{\rm in} -\gamma_{\rm in}'+v)}\right)
\end{equation}
where the constant $v$ is determined by $p=\wp(v)$.

\subsection{The $t$-equation}
The solution procedure of the $t$-equation is similar to the $\phi$-equation. However, in the case of charged particles, some integrals cannot be solved with the current methods.

\subsubsection{Hyperelliptic case}
In the case charged particles $q\neq 0$ we use the substitution $r=\pm \frac{b_5}{4y}+r_0$ and knowing that $\dd\gamma=\frac{y\dd y}{\sqrt{P_5(y)}}$ we can write the $t$-equation \eqref{eqn:t-equation} as
\begin{equation}
 \dd t = \left( A_0 + A_1 y + \frac{A_2}{y-p}  + \frac{B_1}{(y-p)^2} + \frac{B_2}{y}  + \frac{B_3}{y^2}\right) \frac{\dd y}{\sqrt{P_5(y)}}
\end{equation}
with the pole $p=\frac{b_5}{4(r_0-1)}$ and the constants $A_i$ and $B_i$ which arise from a partial fraction decomposition and can be expressed in terms of the parameters of the black hole and the test particle. The terms containing $A_i$ can be integrated as shown in section \ref{sec:phi-solution}. Unfortunately, the terms containing $B_i$ cannot be integrated analytically with the current methods.

\subsubsection{Elliptic case}
In the case  $q=0$ we use the substitutions from section \ref{sec:r-solution}, which transform the $r$-equation \eqref{eqn:r-equation2} into the Weierstra{\ss} form (equation \eqref{eqn:weierstrass-form}). The the $t$-equation \eqref{eqn:t-equation2} can be written as
\begin{equation}
 \dd t = \left( A_0 + \sum _{i=1}^2 \frac{A_i}{y-p_i}  + \sum _{i=1}^2 \frac{B_i}{(y-p_i)^2} \right) \frac{\dd y}{\sqrt{P_3(y)}}
 \label{eqn:tparfrac2}
\end{equation}
with the poles $p_1=\frac{b_2}{12}-\frac{b_3}{4(r_0-1)}$ and $p_2=\frac{b_2}{12}$. The constants $A_i$ and $B_i$ (which are different from the constants in the hyperelliptic case) arise from a partial fraction decomposition and can be expressed in terms of the parameters of the black hole and the test particle. Here it is possible to integrate equation \eqref{eqn:tparfrac2} (compare \cite{Kagramanova:2010bk}, \cite{Grunau:2010gd} and \cite{Willenborg:2018zsv}), so that the solution of the $t$-equation \eqref{eqn:t-equation2} is
 \begin{align}
  t (\gamma) &=t_{\rm in} + A_0 (\gamma -\gamma_{\rm in}) + \sum _{i=1}^2 \frac{A_i}{\wp'(v_i)}\left( 2\zeta(v_i)(\gamma -\gamma_{\rm in}') + \ln\frac{\sigma(\gamma -\gamma_{\rm in}'-v_i)}{\sigma(\gamma_{\rm in} -\gamma_{\rm in}'-v_i)} - \ln\frac{\sigma(\gamma -\gamma_{\rm in}'+v_i)}{\sigma(\gamma_{\rm in} -\gamma_{\rm in}'+v_i)}\right) \nonumber\\
  &- \sum _{i=1}^2 B_i\frac{\wp''(v_i)}{\wp'(v_i)^3} \left( 2\zeta(v_i)(\gamma -\gamma_{\rm in}') + \ln\frac{\sigma(\gamma -\gamma_{\rm in}'-v_i)}{\sigma(\gamma_{\rm in} -\gamma_{\rm in}'-v_i)} - \ln\frac{\sigma(\gamma -\gamma_{\rm in}'+v_i)}{\sigma(\gamma_{\rm in} -\gamma_{\rm in}'+v_i)}\right) \nonumber\\
   &-\sum _{i=1}^2 \frac{B_i}{\wp'(v_i)^2} \left[2 \wp(v_i) (\gamma -\gamma_{\rm in}) + 2 (\zeta(\gamma -\gamma_{\rm in}') - \zeta(\gamma_{\rm in} -\gamma_{\rm in}')) + \frac{\wp'(\gamma -\gamma_{\rm in}')}{\wp(\gamma -\gamma_{\rm in}') - \wp(v_i)} - \frac{\wp'(\gamma_{\rm in} -\gamma_{\rm in}')}{\wp(\gamma_{\rm in} -\gamma_{\rm in}') - \wp(v_i)} \right]
\label{eqn:tsol}
\end{align}
where the constants $v_i$ are determined by $p_i=\wp(v_i)$. Note that there is a printing error concerning the signs in equation (62) and (63) of \cite{Willenborg:2018zsv}, which we corrected in equation \eqref{eqn:tsol}.

\subsection{The orbits}
\label{sec:orbits}

\begin{figure}[p]
	\centering
	\subfigure[Bound orbit with $\delta=1, q=0.7, J=0.28, L=-3.6, E=0.91$.]{
		\includegraphics[width=0.4\textwidth]{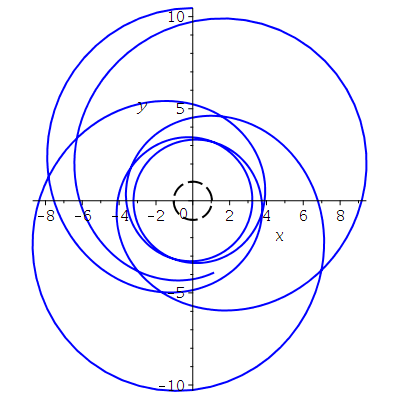}
	}
	\subfigure[Bound orbit with $\delta=1, q=2.5, J=0.6, L=8.8, E=0.955$.]{
		\includegraphics[width=0.4\textwidth]{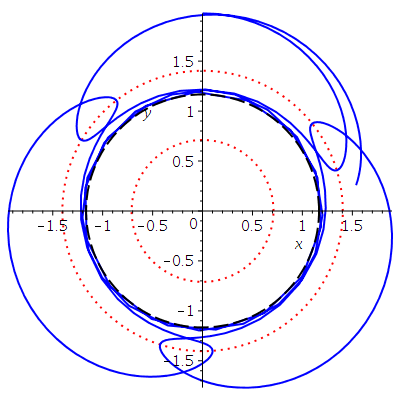}
	}

	\subfigure[Bound orbit behind the horizon with $\delta=1, q=0.1, J=0.4, L=-7, E=6.2$.]{
		\includegraphics[width=0.4\textwidth]{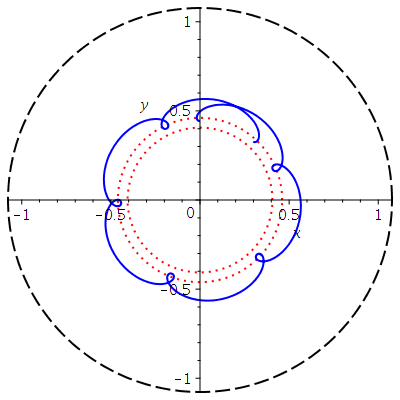}
	}
	\subfigure[Many-world bound orbit with $\delta=1, q=0.8, J=0.34, L=-4.1, E=1.01$.]{
		\includegraphics[width=0.4\textwidth]{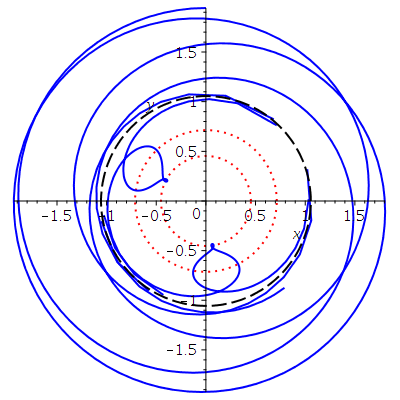}
	}

	\subfigure[Escape orbit with $\delta=1, q=0.1, J=-0.4, L=4, E=1.381$.]{
		\includegraphics[width=0.4\textwidth]{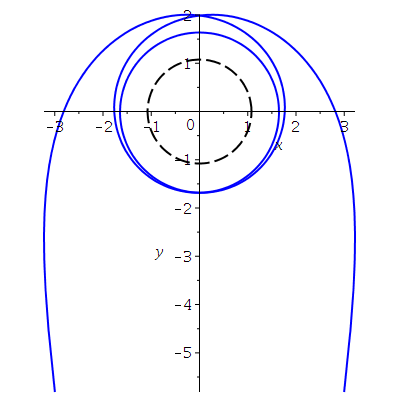}
	}
	\subfigure[Two-world escape orbit with $\delta=1, q=0.7, J=-0.2, L=2.7, E=1.3$]{
		\includegraphics[width=0.4\textwidth]{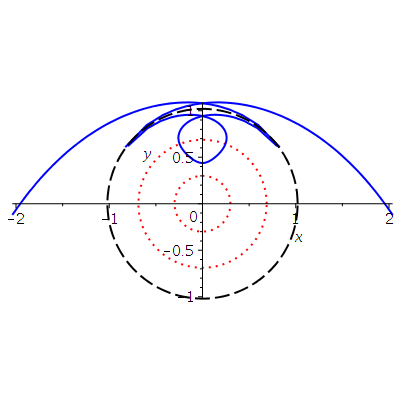}
	}
	\caption{Different orbits of charged particles in the spacetime of an extremal black hole in Kaluza-Klein theory. The blue curves are the orbits, that black dashed circles indicate the position of the horizon and the red dotted circles show the so-called turnaround boundaries.}
 \label{pic:orbits}
\end{figure}

In this section, the analytical solutions are used to plot the orbits of charged particles in the spacetime of an extremal black hole in Kaluza-Klein theory. The orbits are plotted in Boyer-Lindquist coordinates
\begin{align}
 x&= \sqrt{r^2+J^2}\sin\phi \, ,\\
 y&= \sqrt{r^2+J^2}\cos\phi \, .
\end{align}
Figure \ref{pic:orbits} shows some examples of the orbits. Bound orbits can be seen in (a), (b), (c) and (d); whereas escape orbits can be seen in (e) and (f). The bound orbit in (c) exists hidden behind the horizon. (d) shows a many-world bound orbit crossing the horizon multible times. A two-world escape orbit crossing the horizon twice is depicted in (f). The test particles in figure (b), (c), (d) and (f) cross the turnaround boundaries (see section \ref{sec:azimuth}) where they change direction and therefore move on ``loops''.

\section{Conclusion}

In this article we studied the motion of electrically charged particles and light in the spacetime of a rotating dyonic black hole in Kaluza-Klein theory. We focused on the extremal case with a single degenerate horizon. This case is of particular interest, since the angular velocity of the horizon vanishes, but the solution can still have angular momentum.

In section \ref{sec:EQM} we saw that the Hamilton-Jacobi equation for particles moving around the extremal Rasheed black hole separates in three cases:
\begin{enumerate}
 \item Charged particles around a non-rotating Rasheed black hole  with $J=0$.
 \item Uncharged massless particles with $\delta=0$ and $q=0$.
 \item Charged particles in the equatorial plane.
\end{enumerate}
In the present article we studied the third case. For electrically charged particles, the equations of motion are of hyperelliptic type and were solved analytically in terms of the Kleininan $\sigma$ function. For uncharged particles the equations simplify to elliptic type and were solved analytically in terms of the Weierstra{\ss} $\wp$, $\sigma$ and $\zeta$ functions. We analyzed the particle motion with the help of effective potentials and parametric plots and gave a full list of all possible orbit types. Moreover we calculated the photon sphere and considered the causality in the spacetime. Here we saw that CTCs exist behind the horizon.

Depending on the parameters of the black hole and the test particle, the azimuthal equation of motion vanishes at certain radii. At these so-called turnaround boundaries, the test particles change their direction. A similar behaviour is usually seen in an ergosphere, which does not exist in this spacetime. We found that the turnaround boundary can intersect with the effective potentials, so that a test particle is at rest at this point. If this happens at the local minimum of one of the effective potentials, a static ring occurs, where particles are at rest at all times.

For future work it would be interesting to consider the particle motion not only in the equatorial plane but also study the cases of charged particles moving around a non-rotating Rasheed black hole ($J=0$) and uncharged massless particles with $\delta=0$ and $q=0$. Furthermore, one could explore the particle motion in the general non-extreme Rasheed metric. However, this will only be possible with numerical techniques.

\section{Acknowledgements}
We would like to thank Jutta Kunz for fruitful discussions. S.G. gratefully acknowledges support by the DFG (Deutsche Forschungsgemeinschaft/ German Research Foundation) within the Research Training Group 1620 ``Models of Gravity.''

\clearpage

\bibliographystyle{unsrt}

\end{document}